\documentstyle[multicol,aps,epsf]{revtex}
\begin{document}
\draft
\def\l{\lambda}
\def\e{\epsilon}
\def\d{\delta}
\def\half{{1\over2}}
\def\O{{\cal O}}
\def\qc{{q_{\rm c}}}
\def\lc{{\l_{\rm c}}}
\def\av#1{\langle#1\rangle}
\def\a{\alpha}
\def\etal{{\it et al.}}
\def\pc{p_{\rm c}}
\def\df{d_{\rm f}}
\def\K{{\tilde K}}
\def\p{{\tilde p}}
\def\P{{\tilde P}}
\title {Percolation Critical Exponents in Scale-Free Networks}
\author{Reuven~Cohen$^{1}$
\footnote  {{\bf e-mail:} cohenr@shoshi.ph.biu.ac.il}, 
Daniel~ben-Avraham$^2$, and Shlomo Havlin$^1$}
\address{$^1$Minerva Center and Department of Physics, Bar-Ilan
university,
Ramat-Gan, Israel}
\address{$^2$Department of Physics,Clarkson University,
Potsdam NY 13699-5820,
USA}
\maketitle
\begin{abstract}

We study the behavior of scale-free networks, having connectivity
distribution
$P(k)\sim k^{-\lambda}$, close to the percolation threshold. We show that 
for networks with $3<\lambda<4$, known to undergo a transition at a finite
threshold of dilution, the critical exponents are different than the 
expected mean-field values of regular percolation in infinite dimensions.
Networks with $2<\l<3$ possess only a percolative phase.  Nevertheless, we
show that in this case percolation critical exponents are well defined,
near
the limit of extreme dilution (where all sites are removed), and that also
then the exponents bear a strong $\l$-dependence. The regular mean-field
values
are recovered only for $\l>4$.
\end{abstract}
\begin{multicols}{2}

Large networks have been attracting considerable
interest~\cite{bar_rev,dor_rev,bar2,cal,cohen,cohen2}.  A useful
characterization of such networks is their degree distribution, $P(k)$, or
the probability that an arbitrary node be connected to exactly $k$ other
nodes.  Many naturally occurring networks (the Internet, nets of social
contacts, scientific collaborations, ecological nets of predator-pray,
etc.,) exhibit a power-law, or {\it scale-free\/} degree distribution:
\begin{equation}
\label{Pk}
P(k)=c k^{-\lambda},\quad k\geq m\;.     
\end{equation}
where $m$ is the minimal connectivity (usually taken to be $m=1$)
and $c$ is a normalization factor.

Imagine that a large network is diluted, by random removal of
a fraction $p$ of its nodes.  If $p$ is small, one expects that the
network
remains essentially connected: there exists a {\it giant component\/} of
connected nodes that constitutes a finite fraction of the total size of
the
original network.  Above a certain threshold of dilution, $\pc$, the giant
component disappears and the network effectively disintegrates.  This is
the problem of percolation~\cite{stauffer,havlin} as defined on
networks.  (Percolation was originally defined on regular lattices.)  The
giant component corresponds to the {\it infinite incipient cluster\/} or 
{\it spanning cluster} which forms only in the {\it percolating phase}, at
dilutions smaller than the percolation threshold $\pc$.

Percolation in scale-free networks is widely
recognized as a key problem of
interest~\cite{bar_rev,dor_rev,bar2,cal,cohen,cohen2}.  Applications range
from the robustness of communication networks (the Internet, phone
networks) in the face of random failure (removal) of a fraction of their
nodes (e.g., random breakdown of routers), to the efficiency of
inoculation
strategies against the spread of disease, to the ecological impact of the
extinction of some species.

For a random network of arbitrary degree distribution, 
the  condition for the existence of a spanning cluster
is~\cite{cohen,molloy,molloy2,remark}:
\begin{equation}
\label{kappa}
\kappa\equiv{\av{k^2}\over \av{k}}>2.
\end {equation}
Suppose a fraction $p$ of the nodes (and their links) are removed from the
network.  (Alternatively, a fraction $q=1-p$ of the nodes are retained.)
The original degree
distribution, $P(k)$, becomes
\begin{equation}
\label{new_dist}
P'(k)=\sum_{l=k}^\infty P(l)\Big({l\atop k}\Big)(1-p)^{k} p^{l-k}\;.
\end {equation}
In view of this new $P'(k)$, Eq.~(\ref{kappa}) yields the percolation
threshold:
\begin{equation}
\label{perc}
\qc=1-\pc={1 \over \kappa-1}\;,
\end {equation}
where $\kappa$ is computed with respect to the original 
distribution, $P(k)$, before dilution.

Applying the criterion~(\ref{perc}) to scale-free networks one
concludes that for $\l>3$ a phase transition exists at a
finite $\qc$, whereas for $2<\l<3$ the transition takes place only at the 
extreme limit of dilution of $\qc=0$~\cite{cohen,cohen2}.  
Here we concern ourselves with the critical exponents associated with the
percolation transition in scale-free networs of $\l>2$.  We show that
critical exponents are well defined (and we compute their $\l$-dependent
values) even in the non-percolative regime of $2<\l<3$.  In the
percolative
range of $\l>3$, we find $\l$-dependent exponents which differ from the
regular mean-field values of percolation in infinite dimensions.  This
irregularity can be traced to the fat tail of the distribution~(1).  The
regular mean-field exponents are recovered only for $\l>4$.

In~\cite{cal,newman} a generating function is built for the 
connectivity  distribution:
\begin{equation}
\label{G0}
G_0 (x)=\sum_{k=0}^\infty P(k) x^k.
\end {equation}
The probability of reaching a site with connectivity $k$ by following a
specific link is $k P(k)/\av{k}$~\cite{cal,molloy,newman}, and the
corresponding generating function for those probabilities is
\begin{equation}
G_1(x)={{\sum k P(k) x^{k-1}}\over {\sum k P(k)}} =
\frac{d}{dx}G_0(x)/\av{k}\;.
\end {equation}
Let $H_1(x)$ be the generating function for the probability 
of reaching a branch of a given size by following a link.   After a
dilution of a fraction $p$ of the sites, $H_1(x)$ satisfies the 
self-consistent equation 
\begin{equation}
\label{H1}
H_1 (x)=1-q+qx G_1 (H_1 (x))\;.
\end {equation}
Since $G_0(x)$ is the generating function for the connectivity of a site,
the generating function for the probability of a site to belong to an
$n$-site cluster is
\begin{equation}
\label{H0}
H_0 (x)=1-q+qx G_0 (H_1 (x))\;.
\end {equation}
$H_0(1)$ is the probability that a site belongs to a cluster of any {\it
finite\/} size.  Thus, below the percolation transition $H_0(1)=1$, while
above
the transition there is a finite probability that a site belongs to the
infinite
spanning cluster: $P_\infty=1-H_0(1)$. It follows that
\begin{equation}
\label {p_inf_perc}
P_\infty (q)=q(1-\sum_{k=0}^{\infty} P(k) u^k)\;,
\end {equation}
where $u\equiv H_1(1)$ is the smallest positive root of
\begin{equation}
\label{u_perc}
u=1-q+{q\over\av{k}}\sum_{k=0}^{\infty}kP(k)u^{k-1}\;.
\end{equation}
This equation can be solved numerically and the solution may be
substituted
into Eq.~(\ref{p_inf_perc}), yielding the size of the spanning cluster in
a 
network of arbitrary degree distribution, at dilution $q$~\cite{cal}.

We now compute the order parameter critical exponent $\beta$.
Near criticality the probability of belonging to the spanning cluster
behaves as 
$P_\infty\sim(q-\qc)^\beta$. For infinite-dimensional systems (such as
a Cayley tree) it is known that $\beta=1$~\cite{stauffer,havlin,book}.
This regular
mean-field result is not always valid, however, for scale-free networks. 
Eq.~(\ref{p_inf_perc}) has no special behavior at
$q=\qc$; the singular behavior comes from $u$. Also, at
criticality $P_\infty=0$ and Eq.~(\ref{p_inf_perc}) imply that
$u=1$.  We therefore examine 
Eq.~(\ref{u_perc}) for $u=1-\epsilon$ and $q=\qc+\delta$:
\begin {equation}
\label{a0}
1-\epsilon=1-\qc-\delta+{(\qc+\delta)\over\av{k}}\sum_{k=0}^\infty
kP(k)(1-\epsilon)^{k-1}.
\end{equation}
The sum in~(\ref{a0}) has the asymptotic form
\begin{eqnarray}
\label{a1}
&&\sum_{k=0}^{\infty}kP(k)u^{k-1} \sim
 \av{k} - \av{k(k-1)}\e  \nonumber\\ 
&&\ \ + \half\av{k(k-1)(k-2)}\e^2 + \cdots +c\Gamma(2-\l)\e^{\l-2}\;,
\end{eqnarray}
where the highest-order analytic term is $\O(\e^n)$,
$n=\lfloor\l-2\rfloor$.
Using this in Eq.~(\ref{a0}), with
$\qc=1/(\kappa-1)=\av{k}/\av{k(k-1)}$, we get
\begin{eqnarray}
\label{a2}
\frac{\av{k(k-1)}^2}{\av{k}}\d=\half\av{k(k-1)(k-2)}\e + \cdots \nonumber\\
+ c\Gamma(2-\l)\e^{\l-3}\;. 
\end{eqnarray}
The divergence of $\d$ as $\l<3$ confirms the lack of a phase transition
in
that regime.  Thus, limiting ourselves to $\l>3$, and keeping only the
dominant term as $\e\to0$, Eq.~(\ref{a2}) implies
\begin{equation}
\label{a3}
\e\sim\cases{
\bigg({\av{k(k-1)}^2\over c\av{k}\Gamma(2-\l)}\bigg)^{1\over\l-3}
\d^{1\over\l-3}
&${3<\l<4},$\cr {2\av{k(k-1)}^2\over\av{k}\av{k(k-1)(k-2)}}\d &${\l>4}.$ }  
\end{equation}
Returning to $P_{\infty}$, Eq.~(\ref{p_inf_perc}), we see that the
singular
contribution in $\e$ is dominant only for the irrelevant range of
$\l<2$.  For
$\l>3$, we find
$P_{\infty}\sim\qc\av{k}\e\sim(q-\qc)^{\beta}$. 
Comparing this to~(\ref{a3}) we finally obtain
\begin{equation}
\label{beta}
\beta=\cases{
{1\over\l-3} &${3<\l<4},$\cr
1            &${\l>4}. $}
\end{equation}

We see that the order parameter exponent $\beta$ attains its usual
mean-field value
only for $\lambda>4$.   Moreover, for $\l<4$ the percolation transition is
higher than 2nd-order: for $3+\frac{1}{n-1}<\lambda<3+\frac{1}{n-2}$ the
transition is of the $n$th-order.
The result~(\ref{beta}) has been reported before in~\cite{cohen2}, and
also
found independently in a different but related model of virus  
spreading~\cite{vesp,vesp2}. The
existence of an infinite-order phase  transition at $\l=3$ for growing
networks of the Albert-Barab\'asi model, has been
reported elsewhere~\cite{cal2,dor}.   These examples suggest that the
critical exponents are not  model-dependent but depend only on~$\l$.

For networks with $\lambda<3$ the 
transition still exists, though at a vanishing threshold, $\qc=0$. The sum
in Eq.~(\ref{a0}) becomes:
\begin{equation}
\label{sum}
\sum_{k=0}^{\infty}kP(k)u^{k-1} \sim
 \av{k} + c\Gamma(2-\l)\e^{\l-2}\;.
\end{equation}
Using this in conjunction with Eq.~(\ref{u_perc}), and remembering that
here
$\qc=0$ and therefore $q=\d$, leads to
\begin{equation}
\e=\Big({-c\Gamma(2-\l)\over\av{k}}\Big)^{1\over3-\l}\d^{1\over3-\l}\;,
\end{equation}
which implies
\begin{equation}
\label{2beta3}
\beta={1\over 3-\l}\;,\qquad 2<\l<3\;.
\end{equation}
In other words, the transition in $2<\l<3$ is a mirror image of the
transition in $3<\l<4$.  An important difference is that $\qc=0$ is
not $\l$-dependent in $2<\l<3$, and the amplitude of $P_{\infty}$ diverges
as
$\l\to2$ (but remains finite as $\l\to4$). 

In~\cite{newman} it was shown that for a random graph of
arbitrary  degree distribution the finite clusters 
follow the usual scaling form:
\begin {equation}
n_s \sim s^{-\tau}e^{-s/s^*}\;.
\end{equation}
Here $s$ is the cluster size, and $n_s$ is the number of clusters of
size $s$.  At criticality $s^*\sim |q-\qc|^{-\sigma}$ diverges and the 
tail of the  distribution behaves as a power law.  
We now derive the exponent $\tau$.
The probability that a site belongs to an $s$-cluster is $p_s=sn_s\sim
s^{1-\tau}$, and is generated by $H_0$:  
\begin{equation}
H_0(x)=\sum p_s x^s\;.
\end{equation} 
The singular behavior of $H_0(x)$ stems from $H_1(x)$, as can be seen
from Eq.~(\ref{H0}).  $H_1(x)$ itself can be expanded from Eq.~(\ref{H1}),
by using the asymptotic form~(\ref{a1}) of $G_1$.  We let $x=1-\e$, as
before, but work at the critical point, $q=\qc$.   With the notation
$\phi(\e)=1-H_1(1-\epsilon)$, we finally get (note that at criticality
$H_1(1)=1$):
\begin{eqnarray}
\label{phi}
-\phi&&=-\qc+(1-\e)\qc[1-\frac{\phi}{\qc}\nonumber\\
&&+\frac{\av{k(k-1)(k-2)}}{2\av{k}}\phi^2
+\cdots+c\frac{\Gamma(2-\l)}{\av{k}}\phi^{\l-2}]\;.
\end{eqnarray}
{}From this relation we extract the singular behavior of $H_0$:
$\phi\sim\e^y$.  Then, using Tauberian theorems~\cite{weiss} it follows
that
$p_s\sim s^{-1-y}$, hence $\tau=2+y$.

For $\l>4$ the term proportional to $\phi^{\l-2}$ in~(\ref{phi}) may be
neglected.  The linear term
$\e\phi$ may be neglected as well, due to the factor $\e$. This leads to
$\phi\sim\e^{1/2}$ and to the usual mean-field result
\begin{equation}
\tau=\frac{5}{2}\;,\qquad \l>4\;.
\end{equation}
For $\l<4$, the terms proportional to $\e\phi$, $\phi^2$ may be
neglected, leading to $\phi\sim\e^{1/(\l-2)}$ and
\begin{equation}
\label{tau}
\tau=2+\frac{1}{\l-2}=\frac{2\l-3}{\l-2}\;,\qquad 2<\l<4\;.
\end{equation}
Note that for $2<\l<3$ the percolation threshold is strictly $\qc=0$.  In
that case we work at $q=\d$ small but fixed, taking the limit $\d\to0$ at
the very end \cite{rem2}.
For growing networks of the
Albert-Barab\'asi model with $\l=3$, it has been shown that 
$sn_s\propto (s\ln s)^{-2}$~\cite{dor}.  This is consistent with $\tau=3$
plus a logarithmic correction.  Related results for scale free trees have
been presented in~\cite{burda}. 

The critical exponent $\sigma$, for the cutoff cluster size, may be too
derived directly.  Finite-size scaling arguments predict~\cite{stauffer}
that
\begin{equation}
\label{sigma}
\qc(\infty)-\qc(N)\sim N^{-{1\over d\nu}} = N^{-{\sigma\over \tau-1}}\;,
\end{equation}
where $N$ is the number of sites in the network, $\nu$ is the correlation
length critical exponent: $\xi\sim(q-\qc)^{-\nu}$, and $d$ is the
dimensionality of the embedding space.  Using a continuous approximation
of
the distribution~(\ref{Pk}) one obtains~\cite{cohen} 
\begin{equation}
\kappa\approx \biggl({2-\l\over3-\l}\biggr) 
{{K^{3-\l}-m^{3-\l}}\over{K^{2-\l}-m^{2-\l}}}\;,
\end{equation}
where $K\sim N^{1/(\l-1)}$ is the largest site connectivity of the
network.
For $3<\l<4$, this and Eq.~(\ref{perc}) yield
\begin{equation}
\qc(\infty)-\qc(N)\sim \Delta\kappa\sim K^{3-\l}\sim 
N^{3-\l\over\l-1}\;,
\end{equation}
which in conjunction with Eq.~(\ref{sigma}) leads to
\begin{equation}
\label{3sigma4}
\sigma={\l-3\over\l-2}\;,\qquad 3<\l<4\;.
\end{equation}
For $\l>4$ we recover the regular
mean-field result
$\sigma=1/2$.
Note that Eqs.~(\ref{sigma}), (\ref{beta}), (\ref{tau}) are consistent
with the known scaling relation:
$\sigma\beta=\tau-2$~\cite{stauffer,havlin,book}. 
For $2<\l<3$, $\qc(\infty)=0$ and 
$\qc(N)\sim K^{\l-3}\sim N^{(\l-3)/(\l-1)}$. Therefore
\begin{equation}
\label{2sigma3}
\sigma={3-\l\over\l-2}\;,\qquad 2<\l<3\;,
\end{equation}
again consistent with the scaling relation $\sigma\beta=\tau-2$ (cf
Eq.~(\ref{2beta3})).

Through similar scaling relations, any two of the percolation exponents
discussed above determine the remaining (static) percolation exponents.  
Thus, for example, the exponent $\gamma$, governing the average size of
finite clusters: $\av{s}\sim(q-\qc)^{-\gamma}$, follows from the scaling
relation
$\gamma=(3-\tau)/\sigma$.  In view of Eqs.~(\ref{tau}), (\ref{3sigma4}),
(\ref{2sigma3}), we get
\begin{equation}
\label{gamma}
\gamma=\cases{
1          &${\l>3},$\cr
-1          &${2<\l<3}. $}
\end{equation}
The negative value of $\gamma$ in the range $2<\l<3$ is, at first sight,
surprising.  However, recall that for that range only the percolating
phase exists.  At the transition point, $\qc=0$, all clusters have
size {\it zero\/}.  As $q$ increases, finite-size clusters begin to show,
consistent with
$\av{s}\sim (q-\qc)^{-\gamma}\sim q$.

In summary, we have shown that percolation critical exponents in
scale-free
networks bear a strong dependence upon the degree distribution exponent
$\l$. This is true even in the range
$3<\lambda<4$, where the percolation transition occurs at a finite
threshold $\qc$: the regular mean-field behavior of percolation in
infinite
dimensions is recovered only for $\l>4$.  Moreover, critical exponents are
well defined also in the most physically relevant region of $2<\l<3$ (the
case of most naturally occurring networks), despite the lack of a
non-percolating phase.  In this regime too the critical exponents depend
strongly on~$\l$.

An interesting exception is the exponent $\gamma$, which attains constant
values independent of $\l$, Eq.~(\ref{gamma}).  This suggests some
underlying deeper principle which we were not yet able to
uncover.

\acknowledgments
We thank the National Science Foundation for support, under grant
PHY-9820569  (D.b.-A.).

\end{multicols}

\begin{thebibliography}{99}
\bibitem{remark} 
   It follows that for scale-free networks with $m=1$ a spanning
   cluster exists, even before dilution, only for $\l$ smaller than a certain
   critical threshold $\lc$ ($\lc\approx3.5$ if the distribution~(1) is 
   discrete, and $\lc=4$ if it is continuous).  This spurious transition can
   be removed, for example by imposing $m\geq 2$, and does not concern us here.
\bibitem{rem2}
   For the case $2<\l<3$, $\tau$ in Eq. (\ref{tau}) represents the 
   singularity of the distribution of branch sizes. For the distribution of
   cluster sizes in this range one has to consider the singularity of $x$ in
   Eq. (\ref{H0}), leading to $\tau=3$ for this range.
\end{thebibliography}
\end{document}